\documentclass[useAMS,usenatbib]{mn2e}

\usepackage{amssymb,amsmath}
\usepackage{graphicx}
\usepackage{natbib}
\usepackage{aas_macros}

\begin{document}

\title[Low-metallicity star formation: Relative impact of metals and magnetic fields]{Low-metallicity star formation: Relative impact of metals and magnetic fields}

\author[Thomas Peters et al.]{
\parbox[h]{\textwidth}{
Thomas Peters$^1$\thanks{E-Mail: tpeters@physik.uzh.ch},
Dominik R. G. Schleicher$^2$,
Rowan J. Smith$^3$,
Wolfram Schmidt$^2$,
and Ralf S. Klessen$^3$
}\vspace{0.4cm}\\
\parbox{\textwidth}{$^1$Institut f\"{u}r Computergest\"{u}tzte Wissenschaften, Universit\"{a}t Z\"{u}rich,
Winterthurerstrasse 190, CH-8057 Z\"{u}rich, Switzerland\\
$^2$Georg-August-Universit\"{a}t, Institut f\"{u}r Astrophysik, Friedrich-Hund-Platz 1, D-37077 G\"{o}ttingen, Germany\\
$^3$Universit\"{a}t Heidelberg, Zentrum f\"{u}r Astronomie, Institut f\"{u}r Theoretische Astrophysik, Albert-Ueberle-Str. 2, D-69120 Heidelberg, Germany}}

\maketitle

\begin{abstract}
Low-metallicity star formation poses a central problem of cosmology, as it determines the characteristic mass scale and distribution
for the first and second generations of stars forming in our Universe. Here, we present a comprehensive investigation assessing the relative
impact of metals and magnetic fields, which may both be present during low-metallicity star formation. We show that the presence of
magnetic fields generated via the small-scale dynamo stabilises the protostellar disc and provides some degree of support against
fragmentation. In the absence of magnetic fields, the fragmentation timescale in our model decreases by a factor of $\sim10$ at the transition
from $Z=0$ to $Z>0$, with subsequently only a weak dependence on metallicity.
Similarly, the accretion timescale of the cluster
is set by the large-scale dynamics rather than the local thermodynamics. In the presence of magnetic fields, the primordial disc
can become completely stable, therefore forming only one central fragment. At $Z>0$, the number of fragments is somewhat reduced in the presence of magnetic fields, though
the shape of the mass spectrum is not strongly affected in the limits of the statistical uncertainties. The fragmentation
timescale, however, increases by roughly a factor of $3$ in the presence of magnetic fields. Indeed, our results indicate comparable
fragmentation timescales in primordial runs without magnetic fields and $Z>0$ runs with magnetic fields.
\end{abstract}

\section{Introduction}

The formation of stars at low metallicities, and in particular the determination of their characteristic mass scales, is a
central problem in cosmology. Low-metallicity star formation is particularly important during the epoch of reionization, where UV photons
from low-metallicity stars turn the intergalactic medium from a neutral into an ionised state \citep{shapiro94,gnedin00,barkana01,schleicher08}.
If their mass is sufficiently low, they may survive until today and serve as observational probes via near-field cosmology \citep[e.g.][]{clark11a,clark11}. Indeed, a number of
extremely metal poor stars (EMPs) has been discovered in the Milky Way and in nearby dwarf galaxies \citep[e.g.][]{frebel10,salvadori10,caffau13},
including the close-to-primordial star SDSS J1029151+172927 with $Z<10^{-5}$~Z$_\odot$  at the heart of the Lion \citep{caffau11,caffau12}.

From a theoretical point of view, the purely primordial stars were considered to be rather massive, with extreme scenarios reaching several
hundred solar masses \citep{abeletal02,bromm02,yoshidaetal08}. More recent studies have indicated the possibility of
fragmentation \citep[e.g.][]{clarketal08,clark11a,clark11,stacyetal10,greif11,greif12,smith11,latif13}. In addition, radiative feedback appears to set a
characteristic mass scale of $\sim50$~M$_\odot$ \citep{hosokawa11,susa13}. However, these mass scales show significant fluctuations. For
instance, large samples consisting of 100 minihalos may also include stars with up to $1000$~M$_\odot$ \citep{hirano13}, while some halos
with particularly high angular momentum or a large degree of turbulence may host stars with less than $1$~M$_\odot$ \citep{clark11a,stacy14}. In more massive primordial halos exposed
to strong radiation backgrounds, supermassive protostars with up to $10^5$~M$_\odot$ may form \citep{latif13b}.
A low-mass star formation mode can be obtained during the merger of primordial minihalos, as shocks enhance
the electron fraction and the formation of H$_2$ and HD \citep{bovino14b}.
A significant spread of the stellar mass scales thus appears to be present already in the primordial case.

In the presence of heavy elements, the cooling of the gas becomes more efficient, thus decreasing the Jeans mass as well as the accretion rate.
One thus typically expects a decrease in the characteristic mass scales. \citet{bromm03} suggested that such a transition occurs at a
metallicity of $\sim10^{-3}$~$Z_\odot$ to $\sim10^{-4}$~$Z_\odot$, where cooling through carbon and oxygen lines becomes efficient. At the same time, cooling through dust
grains can be important even for dust-to-gas ratios of $\sim10^{-5}$ times the ratio in the solar neighborhood, potentially triggering
fragmentation in close-to-primordial environments \citep{schneider03}. Indeed, this mechanism was potentially important for the formation
of SDSS J1029151+172927, the extremely metal poor star at the heart of the Lion \citep[see e.g.][]{schneider12,klessen12}. The impact of
cooling through metal lines and dust has also been explored in detailed one-zone models by \citet{omukai05,omukai08} and \citet{shrader10}, while
\citet{cazaux04,cazaux09} and \citet{latif12} have demonstrated the importance of H$_2$ formation on dust grains, which can strongly influence
the thermal evolution at metallicities of $\sim10^{-4}$~$Z_\odot$. Detailed three-dimensional (3D) simulations following the impact of dust cooling
during gravitational collapse further have shown that the thermal evolution in 3D can deviate from the results found in one-zone models, in particular
if the collapsing clouds are rotating \citep{dopcke11,dopcke13}. Simulations exploring low-metallicity star formation for metallicities up
to  $10^{-2}$~$Z_\odot$ for number densities till $100\,$cm$^{-3}$ have further been pursued by \citet{jappsen07,jappsen09a,jappsen09b,smith09,hocuk10} and \citet{aykutalp11}.

All in all, the equation of state resulting from the cooling processes discussed here therefore influences the characteristic mass scale of the
clumps \citep{liklessenmaclow03}, as well as the formation of filaments \citep{petersetal12c}. In particular in simulations employing a high
resolution per Jeans length, it is important that not only the cooling processes are accurately employed, but also that high-order numerical
solvers are used when solving the rate equations \citep{bovino13,bovino14}. The publicly available package KROME\footnote{Webpage KROME: http://kromepackage.org/}
for the modeling of chemistry with such high-order solvers was recently released by \citet{grassi14}. While such a detailed modeling is
certainly desirable in the future, we aim here to reproduce only the main features induced by the cooling, and will therefore adopt a
parametrised equation of state derived from the 3D calculations by \citet{dopcke13}. Using a similar technique, \citet{shrader14a} have
recently followed the formation of the first stellar cluster due to metal line cooling, including the formation of a low-mass star \citep{shrader14b}.
The metallicities required for these scenarios can be reached via supernova feedback from previous generations \citep[e.g.][]{greifetal08,greifetal10,ritter12,wiseetal12,seifried14}.

While the metallicity is certainly important in regulating the stellar mass scale, it has been speculated early that also magnetic fields
may have a strong impact on star formation in the high-redshift Universe \citep[e.g.][]{pudritz89,tan04,silk06}. A central question concerns
however their initial field strength, which is highly uncertain in primordial scenarios \citep{grasso01}. Seed fields can be provided
through a number of astrophysical mechanisms \citep[e.g.][]{biermann50,schlickeiser12,schlickeiser13,shiromoto14}. A particularly efficient
amplification mechanism for initially weak seeds is the small-scale dynamo, a process producing strong tangled fields within a few eddy-turnover
times \citep{schekochihin02,brandenburg05,federrath11b,schober12b,schleicher13}. This process was therefore suggested to provide strong
magnetic fields during the formation of the first stars and galaxies \citep{arshakian09,schleicher10c,sur10,souza10,schober12,sur12,turk12,latif13c}.
In the case of large-scale coherent magnetic fields, \citet{machida06,machida08} have shown that they can lead to the suppression of fragmentation
and the formation of the first jets in the Universe. More recently, they have provided an in-depth investigation regarding the interplay of the
magnetic field with fragmentation and the transport of angular momentum \citep{machida13}, while \citet{latif13d} have shown that magnetic fields
generated via the small-scale dynamo can help to suppress fragmentation during the formation of supermassive black holes.

So far, there is however no study concerning the impact of magnetic fields for low-metallicity star formation, i.e. with metallicities above zero.
Here, we provide the first exploration of the combined impact of cooling through metals and dust in the presence of strong tangled
magnetic fields, as provided via the small-scale dynamo. Our simulation setup is presented in section 2, while the results are analysed and described
in section 3. A summary and discussion is provided in section 4.

\section{Simulations}

We present three-dimensional magnetohydrodynamical (MHD) collapse simulations, using the adaptive-mesh code FLASH \citep{fryxell00} and
an MHD solver that preserves positive states \citep{bouchut07,waagan09}.
The initial conditions for our simulations are similar to the ones in \citet{petersetal12c}. We use a minihalo from a cosmological simulation
of \citet{greif11}. This halo corresponds to Halo~4 in their nomenclature and has a virial mass of $3.1 \times 10^5\,$M$_\odot$ and a virial
radius of $97\,$pc. Once the halo has collapsed to a central density of $2.8 \times 10^{-15}\,$g\,cm$^{-3}$, we cut out the central $8000$~AU of the simulation.
This leaves us with a rapidly collapsing sphere of gas containing $101.3\,$M$_\odot$ and a cosmologically consistent density and velocity
structure, which we can use for our initial condition.
We use outflow boundary conditions for the hydrodynamics and isolated boundary conditions for the gravity solver. Since the free-fall time
of the gas at the boundary of the simulation box is more than three times longer than our maximum simulation run-time, we do not expect
this cut-out technique to affect our results much. In fact, at the end of our simulations the gas at the boundaries has barely started
to collapse.
The initial condition for the magnetic
field is identical to \citet{petersetal12c}, except for the normalization. The magnetic field has a power-law spectrum $P_B(k)\propto k^{3/2}$  on large
scales \citep{kazantsev68}, peaks on a scale of $1250\,$AU, corresponding to 20 grid cells in our initial setup, and drops with $P_B(k)\propto k^{-4}$ on smaller scales. This magnetic field spectrum
represents an idealization of the spectra measured in collapse simulations by \citet{federrath11}.

\begin{figure}
\centerline{\includegraphics[height=170pt]{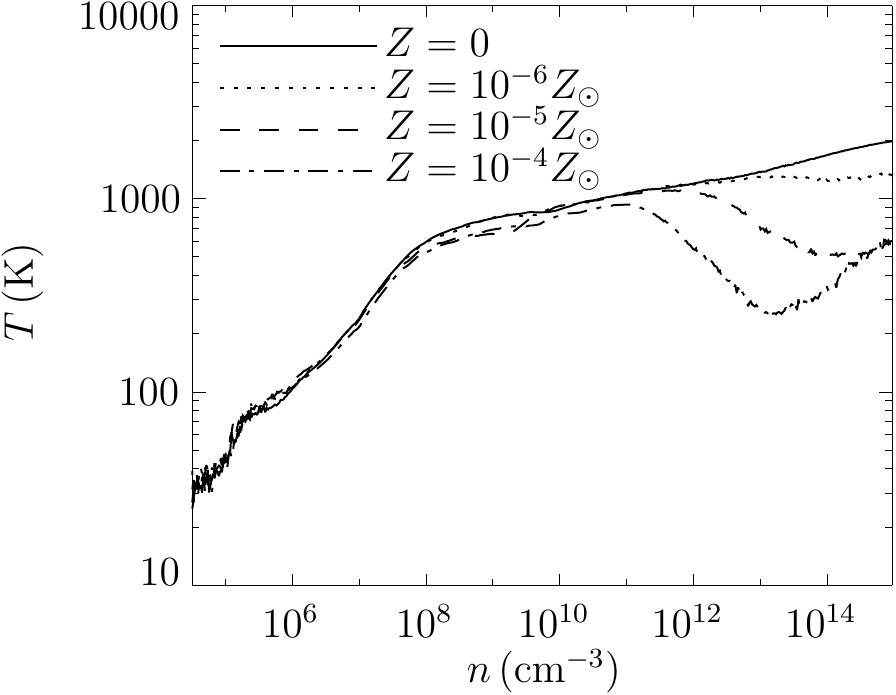}}
\caption{Temperature-density phase diagram for the different metallicities $Z = 0, 10^{-6} Z_\odot, 10^{-5} Z_\odot$ and $10^{-4} Z_\odot$.}
\label{fig:eos}
\end{figure}

The temperature field is set by a barotropic equation of state instead of the polytropic equation of state employed by \citet{petersetal12c}. 
We use a look-up table generated from the simulations by \citet{dopcke13},
which include a time-dependent chemical network to model the thermodynamics of a mixture of low-metallicity gas and dust. Figure~\ref{fig:eos} shows
the temperature-density phase diagram for the different metallicities $Z = 0, 10^{-6} Z_\odot, 10^{-5} Z_\odot$ and $10^{-4} Z_\odot$.

For each metallicity, we have run simulations with an initial magnetic field strength of $B_0 = 0$
(Z0B0, Z6B0, Z5B0 and Z4B0 for a metallcity of $Z = 0, 10^{-6} Z_\odot, 10^{-5} Z_\odot$ and $10^{-4} Z_\odot$, respectively)
and $B_0 = 10^{-2}\,$G (Z0B2, Z6B2, Z5B2 and Z4B2). For $Z = 0$, we have run an additional simulation with $B_0 = 3 \times 10^{-3}\,$G (Z0B3).
For a summary of the main model parameters, see Table~\ref{table:sim}. These
values have been adopted in order to have a reference case corresponding to the absence of a magnetic field, as well as simulations which start already
close to saturation, i.e. where turbulent and magnetic energies are comparable. Such a state is indeed expected due to magnetic field amplification
via the small-scale dynamo \citep{schleicher10c,schober12}. In the simulations with $B_0 = 10^{-2}\,$G ($B_0 = 3 \times 10^{-3}\,$G), the initial magnetic
energy amounts to 45\% (4\%) of the thermal energy, 11\% (2\%) of the gravitational energy and 114\% (10\%) of the kinetic energy of the halo.
The simulations with $B_0 = 10^{-2}\,$G therefore represent an extreme case where the magnetic field has been maximally amplified.

Since we want to follow all simulations to a similarly high density, the different temperatures at a given density for the various metallicities
result in different sizes of the Jeans length and Jeans mass at the resolution limit. We introduce sink particles \citep{federrathetal10}
at a threshold density $n_\mathrm{thres} = 10^{15}\,$cm$^{-3}$ and set the sink accretion radius to half the Jeans length at this density.
The adaptive mesh refinement is set up such that it always resolves the Jeans length during the collapse with at least 32 grid cells and that
it resolves the sink particle radius with at least 4 grid cells. Table~\ref{table:sim} summarises the parameters and resolution limits of
the simulations. We have stopped the simulations when the total cluster mass reached $3.75\,M_\odot$.

\begin{table*}
\caption{Simulations}
\label{table:sim}
\begin{center}
\begin{tabular}{lrccccccccc}
\hline
Name & \multicolumn{1}{c}{$Z$} & \multicolumn{1}{c}{$B_0$} & \multicolumn{1}{c}{$r_\mathrm{acc}$}& \multicolumn{1}{c}{$M_\mathrm{Jeans}$} & $N$  & \multicolumn{1}{c}{$M_\mathrm{mean}$} & \multicolumn{1}{c}{$M_\mathrm{median}$}
& \multicolumn{1}{c}{$\sigma_M^2$} & \multicolumn{1}{c}{$t_\mathrm{first}$} & \multicolumn{1}{c}{$\dot{M}_\mathrm{mean}$} \\
& \multicolumn{1}{c}{($Z_\odot$)} & \multicolumn{1}{c}{(G)} & \multicolumn{1}{c}{(AU)} & \multicolumn{1}{c}{($10^{-2} M_\odot$)} & & \multicolumn{1}{c}{($M_\odot$)} &\multicolumn{1}{c}{($M_\odot$)} &
\multicolumn{1}{c}{($M_\odot$)} & \multicolumn{1}{c}{(kyr)}  & \multicolumn{1}{c}{($10^{-3} M_\odot$\,yr$^{-1}$)} \\ \hline
Z0B0 & $0$ & $0$              & 1.63 & 7.12 & 4 & 0.94 & 1.13 & 0.20 & 1.84 & 10.1 \\
Z0B3 & $0$ & $3 \times 10^{-3}$ & 1.63 & 7.12 & 3 & 1.25 & 1.15 & 0.85 & 1.83 & 8.51 \\
Z0B2 & $0$ & $10^{-2}$         & 1.63 & 7.12 & 1 & 3.75 & --- & --- & 1.83 & 14.6 \\
\hline
Z6B0 & $10^{-6}$ & $0$         & 1.31 & 3.73 & 11 & 0.34 & 0.21 & 0.10 & 1.84 & 4.42 \\
Z6B2 & $10^{-6}$ & $10^{-2}$    & 1.31 & 3.73 & 6  & 0.63 & 0.71 & 0.07 & 1.83 & 6.77 \\
\hline
Z5B0 & $10^{-5}$ & $0$         & 0.89 & 1.17 & 13 & 0.29 & 0.17 & 0.08 & 1.72 & 3.90 \\
Z5B2 & $10^{-5}$ & $10^{-2}$    & 0.89 & 1.17 & 11 & 0.34 & 0.37 & 0.11 & 1.70 & 3.37 \\
\hline
Z4B0 & $10^{-4}$ & $0$         & 0.93 & 1.32 & 19 & 0.20 & 0.07 & 0.06 & 1.70 & 2.07 \\
Z4B2 & $10^{-4}$ & $10^{-2}$    & 0.93 & 1.32 & 11 & 0.34 & 0.32 & 0.11 & 1.69 & 2.87 \\
\hline
\end{tabular}
\medskip\\
Metallicity $Z$, initial magnetic field $B_0$, sink accretion radius $r_\mathrm{acc}$ and Jeans mass at the resolution limit
$M_\mathrm{Jeans}$ for the different simulations as well as the number of stars $N$, the mean mass $M_\mathrm{mean}$, median mass $M_\mathrm{median}$
and the variance of the mass spectrum $\sigma_M^2$ after approximately $3.75\,M_\odot$ have been accreted onto stars, the time
at which the first sink particle forms measured from the beginning of the simulation $t_\mathrm{first}$ and the average
sink accretion rate over the simulation run-time $\dot{M}_\mathrm{mean}$.
\end{center}
\end{table*}

\section{Analysis}

\subsection{Magnetic Field Morphology and Filament Formation}

\begin{figure*}
\centerline{\includegraphics[width=350pt]{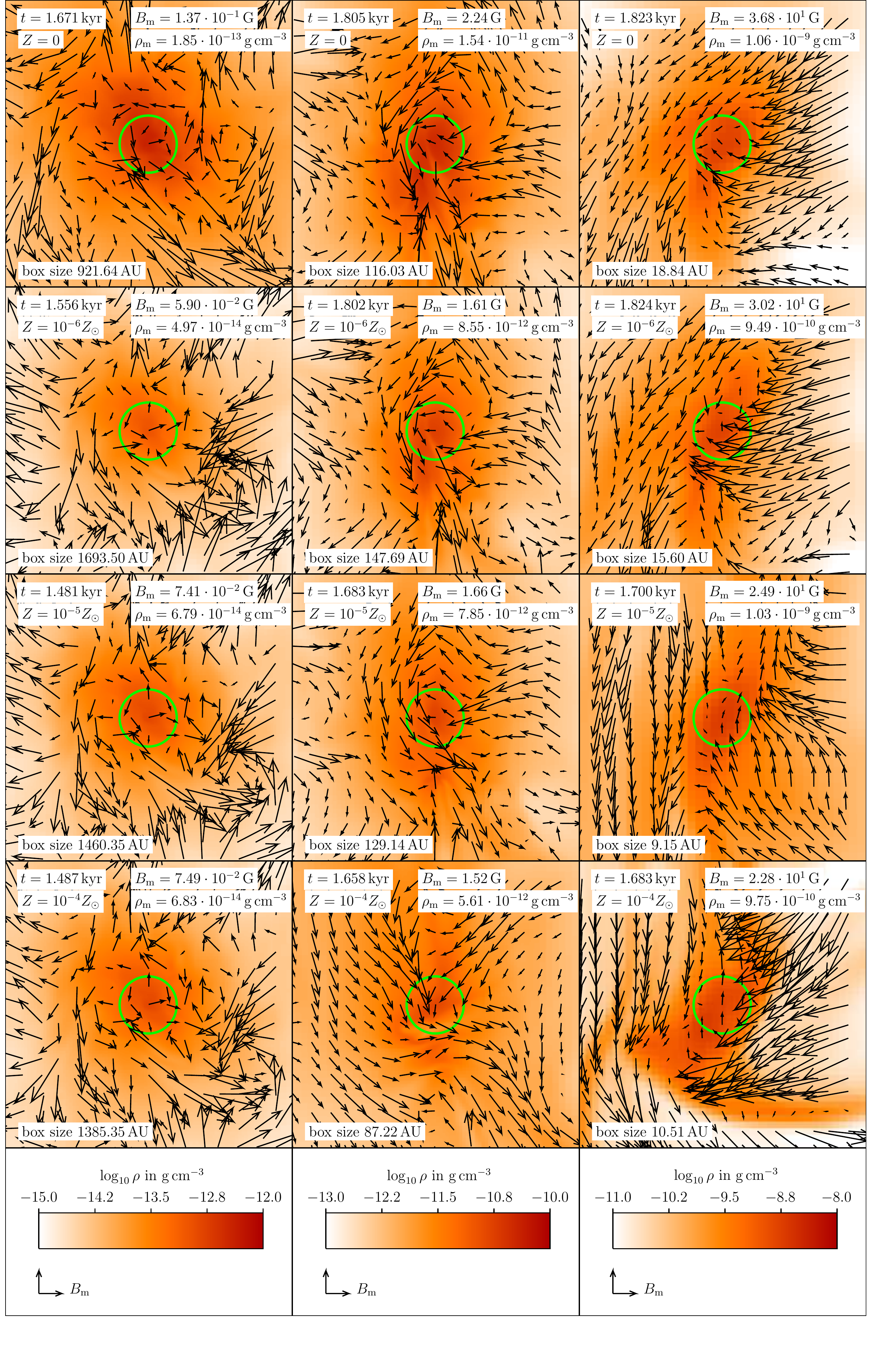}}
\vspace{5pt}
\caption{Magnetic field and density structure for the runs Z0B2, Z6B2, Z5B2 and Z4B2 as a function of time. Rows are different metallicities
($Z = 0, 10^{-6}, 10^{-5}, 10^{-4} Z_\odot$, from top to bottom), columns are different times (time advances from left to right). The mean
magnetic field $B_\mathrm{m}$ and mean density $\rho_\mathrm{m}$ within the Jeans volume are indicated in each panel.
The magnetic field vectors have been rescaled for plotting by $(\rho/\rho_\mathrm{m})^{2/3}$, and a field strength of $B_\mathrm{m}$ at a density
of $\rho_\mathrm{m}$ corresponds to an arrow of the length given in the legend. The green circle marks the Jeans volume.}
\label{fig:magneticfield}
\end{figure*}

Figure~\ref{fig:magneticfield} shows density slices and magnetic field vectors at three different times during the collapse prior
to sink particle formation for the magnetic runs Z0B2, Z6B2, Z5B2 and Z4B2. The figure shows that the magnetic field morphology
and the formation of filaments by turbulence during the collapse depends critically on the thermodynamics. In agreement with
our findings from more idealised simulations with a constant polytropic exponent \citep{petersetal12c}, we observe a virialised,
central core and tangled magnetic field vectors as long as the equation of state is super-isothermal,
but notice strong shocks with sharp density contrasts and coherent magnetic field vectors on scales much larger than the Jeans
volume during isothermal or even sub-isothermal collapse phases. Sub-isothermal collapse occurs in simulations with $Z = 10^{-5} Z_\odot$
and $Z = 10^{-4} Z_\odot$ at densities above $n \gtrsim 10^{12}\,$cm$^{-3}$ and $n \gtrsim 10^{11}\,$cm$^{-3}$, respectively.
The simulations with $Z = 0$ and $Z = 10^{-6} Z_\odot$ are only slightly super-isothermal for $n \gtrsim 10^8\,$cm$^{-3}$, and
so the differences in the qualitative behaviour of the simulations becomes smaller as the collapse proceeds.

\subsection{Sink Particle Formation}

\begin{figure*}
\centerline{\includegraphics[width=350pt]{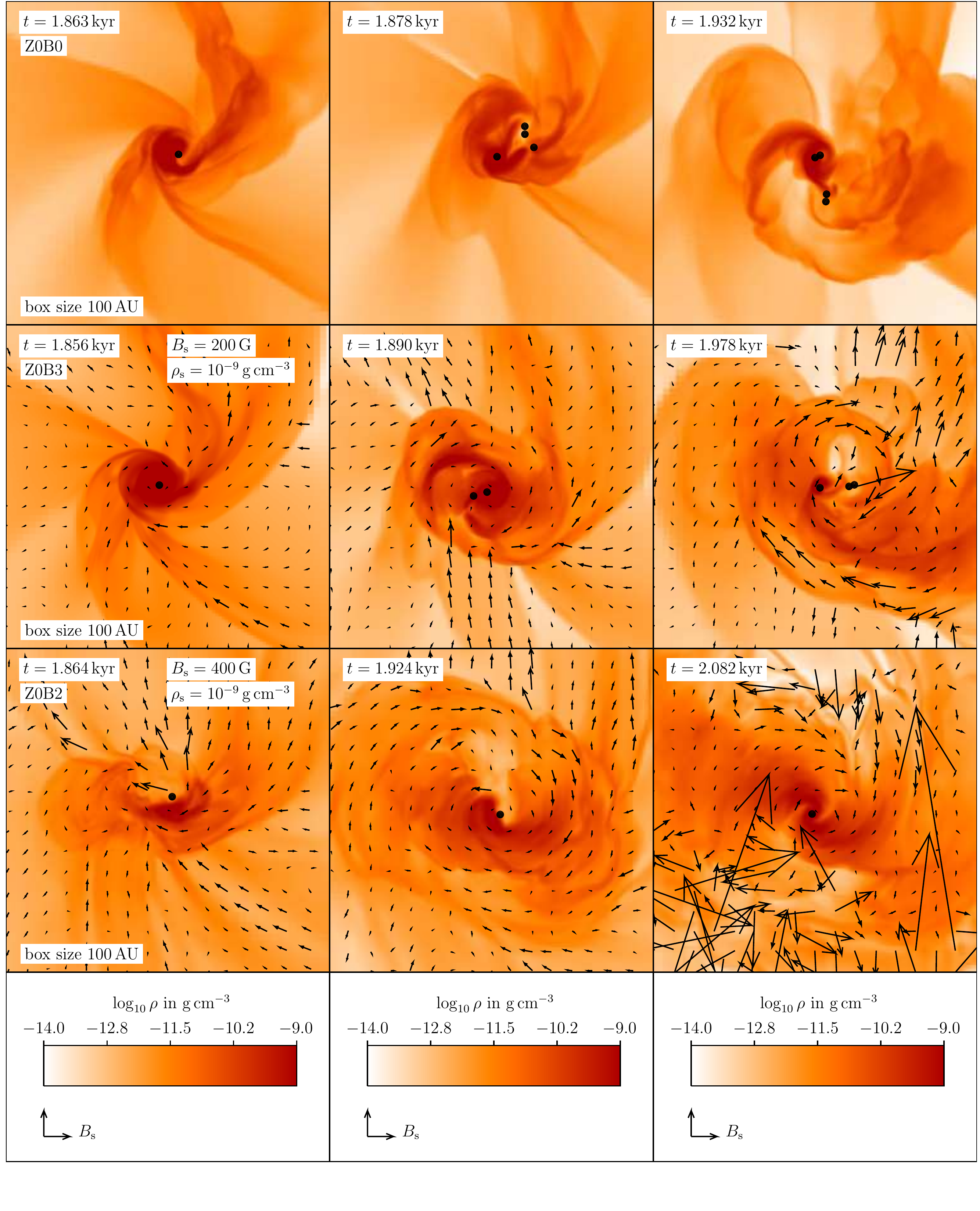}}
\vspace{5pt}
\caption{Magnetic field and density structure for the runs Z0B0, Z0B3 and Z0B2 as a function of time. Rows are different initial magnetic field strengths
($B_0 = 0, 3 \cdot 10^{-3}, 10^{-2}\,$G, from top to bottom), columns are different times (time advances from left to right). The snapshots show
the simulations when the cluster masses have reached $1$, $2$ and $3.75\,M_\odot$, respectively.
The magnetic field vectors have been rescaled for plotting by $(\rho/\rho_\mathrm{s})^{2/3}$, and a field strength of $B_\mathrm{s}$ at a density
of $\rho_\mathrm{s}$ corresponds to an arrow of the length given in the legend. Black dots represent sink particles.}
\label{fig:Z0}
\end{figure*}

Figure \ref{fig:Z0} shows density slices and magnetic field vectors for the three simulations with $Z = 0$. The runs
with $Z > 0$ behave similarly and are displayed in Appendix~\ref{app:magden} for the sake of completeness.
In all cases, sink particles form in disc-like, rotating
structures at the densest parts of filamentary density enhancements. These flattened, rotationally supported structures are
called pseudo-discs \citep{gallishu93a,gallishu93b}. Their velocity fields can deviate significantly from Keplerian profiles
as a result of strong gravitational instability \citep[e.g.][]{petersetal11a}.
For $Z \leq 10^{-6} Z_\odot$, these pseudo-discs appear to be
more stable to fragmentation compared to the simulations with $Z \geq 10^{-5} Z_\odot$, where sink particle formation and shocks destroy
the central, dense structure and sink particle formation proceeds along the forming filaments.
We stress that in the following, we will use the term ``disc'' as abbreviation of ``rotationally-flattened structure''.
Our discs should not be confused with thin discs, which are dominated by the gravity field of the star at the centre, and only
gradually become unstable as more gas falls onto the disc and cannot be transported inwards fast enough. We neither
see coherently rotating velocity fields nor pronounced radial transport in these discs.

We note that, since we scale the magnetic field vectors by $(\rho/\rho_\mathrm{s})^{2/3}$
with a reference density $\rho_\mathrm{s}$ to visualise the magnetic field over several orders of magnitude in density, very long arrows can originate from
areas with a density much lower than $\rho_\mathrm{s}$ if the magnetic field is comparatively strong there. Therefore, for fixed $\rho_\mathrm{s}$, larger
arrows mean greater magnetic field amplification beyond pure flux freezing.

Gravoturbulent fragmentation at the sites of sink particle formation is so strong that a coherent disc magnetic field cannot build up. The only exception
are runs Z0B3 and Z0B2, in which the most pronounced disc structures form. We have followed the evolution of these disks for roughly 5 and 10
orbital times, respectively. Nevertheless, we found no evidence for magnetically-driven
outflows in our simulations. This is probably because we would need to simulate more orbital times to allow a strong
toroidal magnetic field to build up. 

It is obvious that the magnetic field reduces fragmentation and sink particle formation for all metallicities (compare Table~\ref{table:sim}).
The extreme case is run Z0B2, in which only a single fragment forms in the centre of the disc during the simulation run-time.
Hence, although our sink particles do not represent finished stars, we expect the magnetic fields to shift the primordial stellar mass
spectrum towards higher masses, similarly to the situation in present-day star formation
\citep[e.g.][]{wangetal10,petersetal11a,hennetal11,comm11,myersetal13}.

Table~\ref{table:sim} displays some basic statistical information about the sink particles that form in the different simulations.
We show the total number of stars $N$ that form during the simulation as well as the mean mass $M_\mathrm{mean}$, median mass $M_\mathrm{median}$
and the variance of the mass spectrum $\sigma_M^2$ when the simulations are stopped. Furthermore, we indicate the time $t_\mathrm{first}$
when the first sink particle forms and the average sink accretion rate $\dot{M}_\mathrm{mean}$, which is defined as the total
mass in sinks at the end of the simulation divided by the cluster age and the number of sink particles.

\subsection{Support Functions}
\label{sec:supfun}

We can quantify the relative importance of thermal, turbulent and magnetic support against
gravitational collapse using analytical methods introduced by \citet{schmidt13}. For a gas with thermal pressure $P$, density $\rho$, velocity $\bmath{v}$
and magnetic field $\bmath{B}$, we define
the thermal support function
\begin{equation}
\Lambda_\mathrm{therm} = -\frac{1}{\rho} \frac{\partial^2 P}{\partial x_i \partial x_i} + \frac{1}{\rho^2} \frac{\partial \rho}{\partial x_i} \frac{\partial P}{\partial x_i},
\end{equation}
the turbulent support function
\begin{equation}
\Lambda_\mathrm{turb} = \frac{1}{2} \left(\omega_i \omega_i - \sqrt{2 S_{ij} S_{ij}}\right)
\end{equation}
with the vorticity $\bmath{\omega} = \nabla \times \bmath{v}$ and rate-of-strain tensor
\begin{equation}
S_{ij} = \frac{1}{2} \left(\frac{\partial v_i}{\partial x_j} + \frac{\partial v_j}{\partial x_i}\right),
\end{equation}
and the magnetic support function
\begin{equation}
\begin{split}
\Lambda_\mathrm{magn} &= \frac{1}{4 \pi \rho} \left[- \frac{\partial^2}{\partial x_i \partial x_i} \left(\frac{1}{2} B^2\right)
+ \frac{\partial B_i}{\partial x_j} \frac{\partial B_j}{\partial x_i} \right] \\
&+ \frac{1}{4 \pi \rho^2} \frac{\partial \rho}{\partial x_i} \left[\frac{\partial}{\partial x_i} \left(\frac{1}{2} B^2\right) - B_j \frac{\partial B_i}{\partial x_j}\right].
\end{split}
\end{equation}
Here we have implicitly assumed summation over repeated indices.
Positive values of $\Lambda_\mathrm{therm}$, $\Lambda_\mathrm{turb}$ and $\Lambda_\mathrm{magn}$ indicate support against gravitational collapse, whereas negative
values mean that collapse is promoted. As a measure of the contribution to the (positive or negative) support against gravity, we use the ratio of these
support functions and the gravitational compression rate $4 \pi G \rho$. Thus, we consider the quantities $\Xi_\mathrm{therm} = \Lambda_\mathrm{therm} / 4 \pi G \rho$,
$\Xi_\mathrm{turb} = \Lambda_\mathrm{turb} / 4 \pi G \rho$ and $\Xi_\mathrm{magn} = \Lambda_\mathrm{magn} / 4 \pi G \rho$. We compute $\Xi_\mathrm{therm}$, 
$\Xi_\mathrm{turb}$ and $\Xi_\mathrm{magn}$ for all grid cells and then derive mass-weighted averages for 50 density bins.

\begin{figure*}
\centerline{\includegraphics[height=140pt]{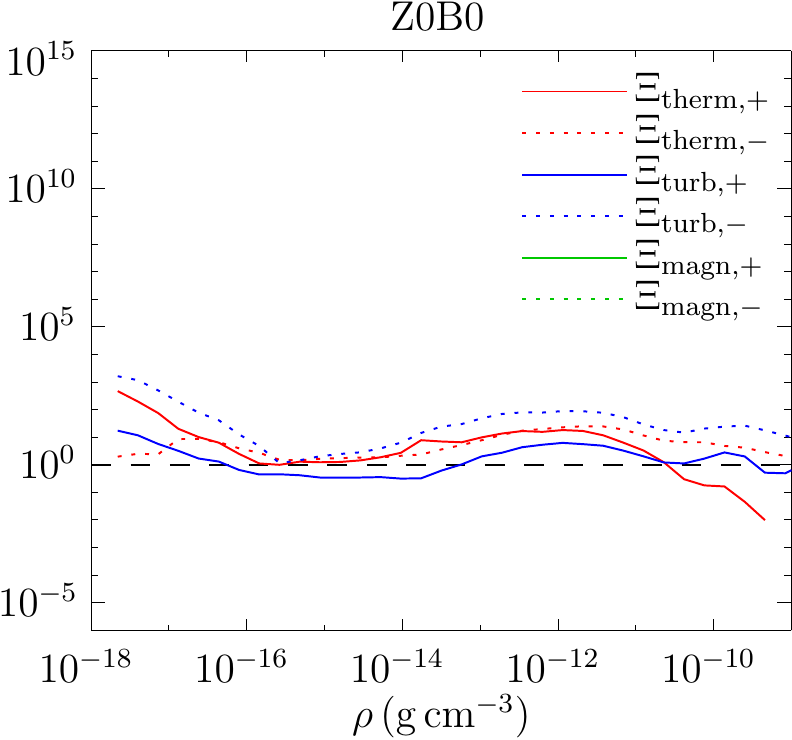}
\includegraphics[height=140pt]{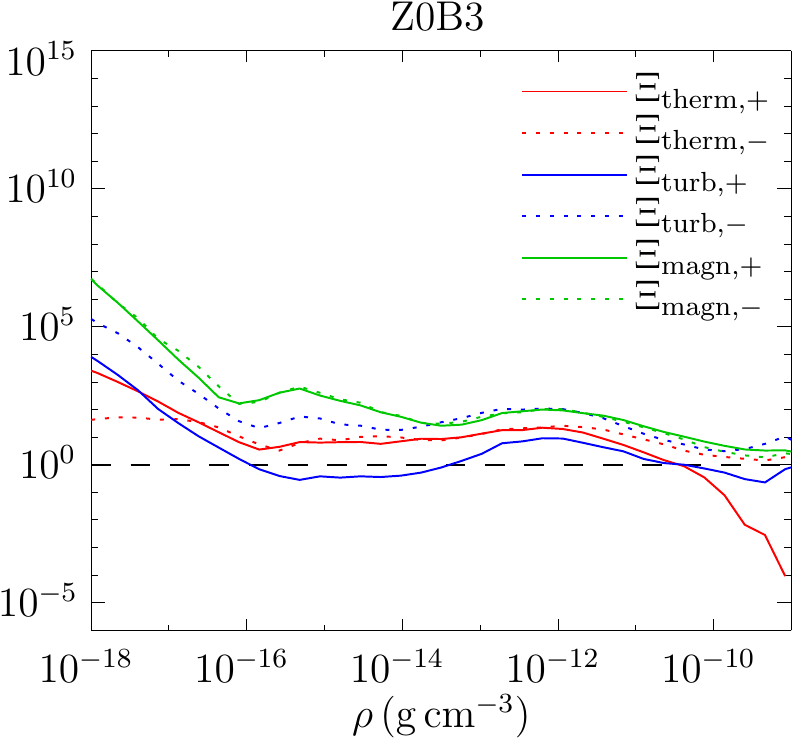}
\includegraphics[height=140pt]{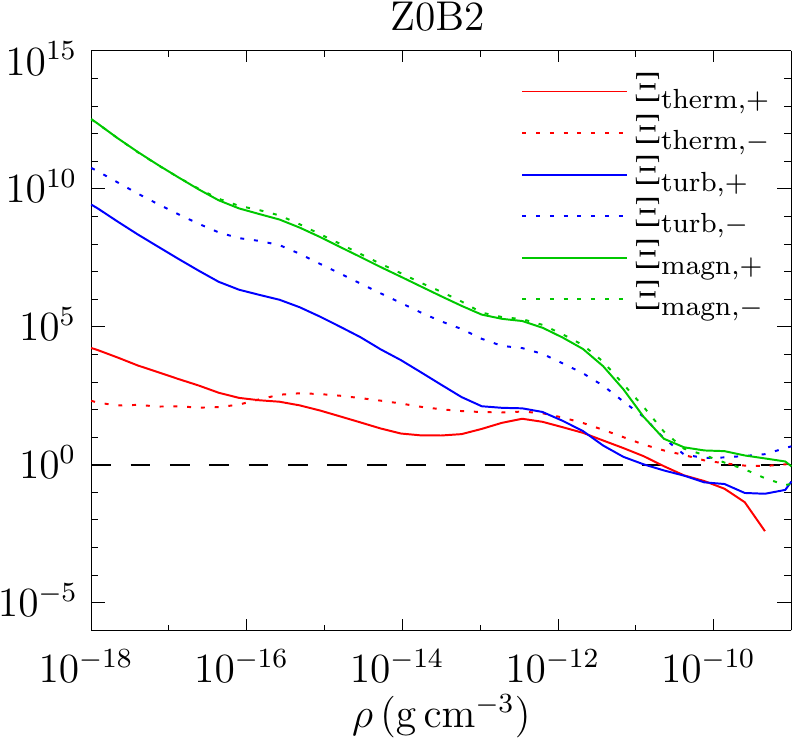}}
\caption{Support functions as a function of density after accretion of $3.75\,M_\odot$ for runs Z0B0, Z0B3 and Z0B2.}
\label{fig:lambdaZ0B0}
\end{figure*}

Figure \ref{fig:lambdaZ0B0} shows $\Xi_\mathrm{therm}$, $\Xi_\mathrm{turb}$ and $\Xi_\mathrm{magn}$
as a function of density for the runs with $Z = 0$. The corresponding plots for $Z > 0$ are deferred to Appendix~\ref{app:supp}. Positive and negative values are plotted
separately. Especially in run Z0B2, the presence of the magnetic field dramatically enhances not only the magnetic support function,
but also the turbulent support function. In the simulations without magnetic fields, the strongest positive support is due to the thermal pressure,
while turbulence provides a predominantly negative support due to compressive motions (see similar results by \citet{latif13e} and \citet{latif13d} for a
primordial collapse in the presence of strong radiative backgrounds).

In the presence of a magnetic field, the positive and negative contributions are nearly equal over a wide range
of densities, with the exception of the highest densities. In contrast, \citet{schmidt13} find pronounced positive
support by magnetic fields. However, there are important differences compared to the scenario we consider here.
Firstly, \citet{schmidt13} compute the support for turbulence produced by external forcing in a periodic box, where
gravitational collapse is triggered by supersonic turbulent compressions of the gas. Secondly, 
no sink particles are inserted in their simulations. As a result, the magnetic field is squeezed into collapsing gas of arbitrarily high density.
In our simulations, on the other hand, the magnetic field is decoupled from the collapsing gas once sink particles are inserted. This limits
somewhat the maximal magnetic pressure that can build up against gravity. Thirdly, we expect our collapsing halo to have a stronger
tendency to produce disc-like structures in comparison to turbulence produced by random forcing, which, averaged over the box, induces
zero angular momentum. These rotating discs might affect the field morphology dramatically by winding up magnetic field lines and building up
magnetic pressure. Nevertheless, we see that the net magnetic support becomes positive at densities of
$10^{-9}-10^{-10}$~g~cm$^{-3}$, where a protostellar accretion disc has formed. This effect is particularly
pronounced in the simulation Z0B2, but also visible in Z0B3 with a weaker magnetic field, as well as in the higher-metallicity simulations. Within the
disc, the magnetic field therefore provides a stabilising contribution, as also reported by \citet{latif13d}. We therefore expect that the fragmentation
timescale increases in the presence of a magnetic field, as supported by a more detailed analysis of the relevant timescales in the system (Section~\ref{sec:timan}).

\subsection{Angular Momentum}

\begin{figure}
\centerline{\includegraphics[height=140pt]{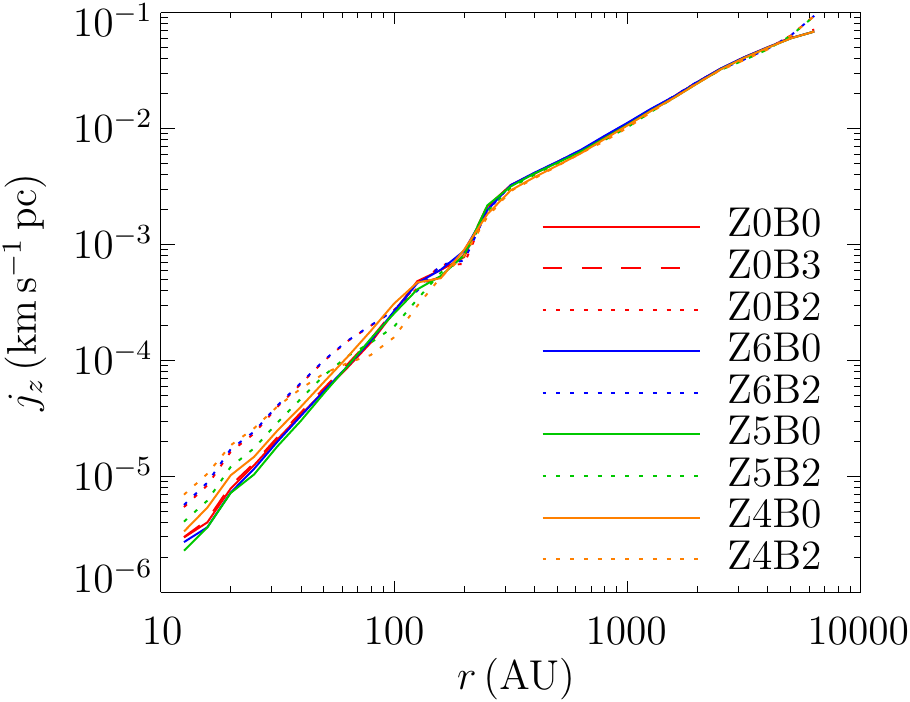}}
\centerline{\includegraphics[height=140pt]{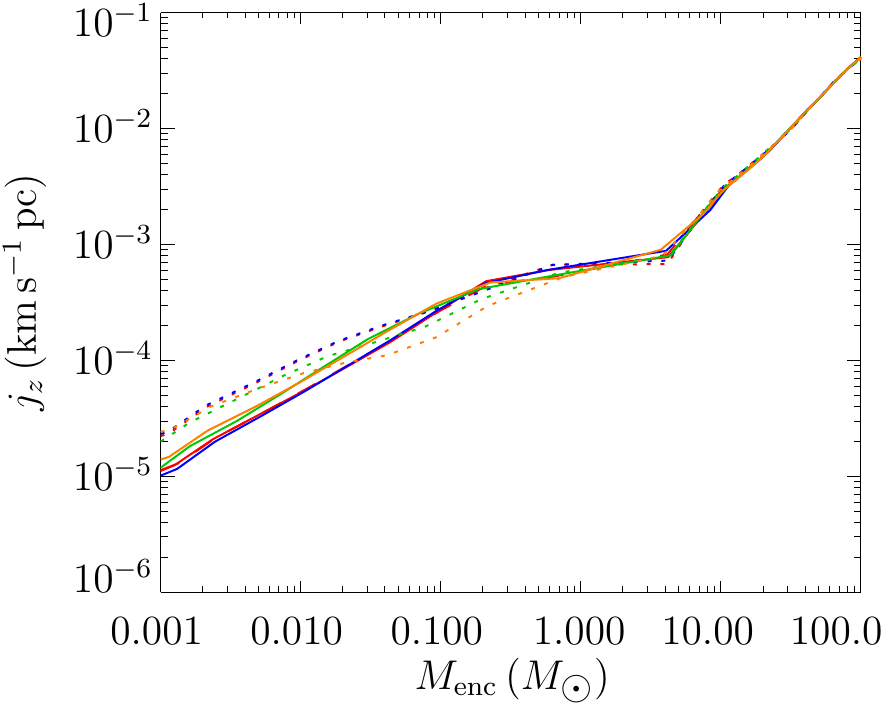}}
\caption{Specific angular momentum $j_z$ for all simulations at the time of first sink formation as function of radius $r$ (\emph{top})
and enclosed mass $M_\mathrm{enc}$ (\emph{bottom}).}
\label{fig:jz}
\end{figure}

Figure~\ref{fig:jz} displays the specific angular momentum $j_z$ at the time of first sink formation as a function of radius and enclosed mass for
all simulations. The values shown are the mass-weighted averages over spherical shells centred on the densest grid cell in the
domain. Differences between the profiles are very small in general, so that even a magnetic field close to
saturation does not seem to substantially impact the specific angular momentum profile.
The magnetised halos have only a slightly larger $j_z$ in the inner $100\,$AU. This behaviour is to be expected for the runs with $Z = 0$,
because here the disc structure, which is aligned approximately orthogonally to the $z$-axis of the simulation grid, becomes much more pronounced when the magnetic
field increases, but for $Z > 0$ it is less clear from the density structures how an enhanced $j_z$ can be interpreted.
The other components of the specific angular momentum vector, $j_x$ and $j_y$, have more complicated, non-monotonic profiles, and
show no strong trend with metallicity or magnetic field strength. In particular, both $j_x$ and $j_y$ change signs several times
as a function of radius, so that the situation gets very difficult to visualise.

In their collapse simulations, \citet{machida13} find that magnetic braking transports angular momentum very efficiently
and prevents the formation of a disc when $B \gtrsim 10^{-12} (n / 1\,$cm$^{-3})^{-2/3}\,$G, which is of a similar magnitude as our
initial field strengths. This difference between the two sets of simulations is most likely caused by the differing initial conditions.
\citet{machida13} initialise their simulations with large-scale ordered magnetic field configurations, whereas our
simulations start from magnetic fields with no global order.

\subsection{Mass Spectra}

Table~\ref{table:sim} summarises the properties of the sink particle mass spectra for the different simulations.
In our simulations, the sink particles represent dense collapsing fragments of gas. Such fragments are the very earliest stages of protostars
as they have collapsed to high densities but have not yet started to contract to the main sequence.
The magnetic field reduces the number of sinks and increases the mean sink mass in all cases. 
For a fixed magnetic field strength, more sink particles form at a given star formation efficiency with growing metallicity.

\begin{figure}
\centerline{\includegraphics[height=160pt]{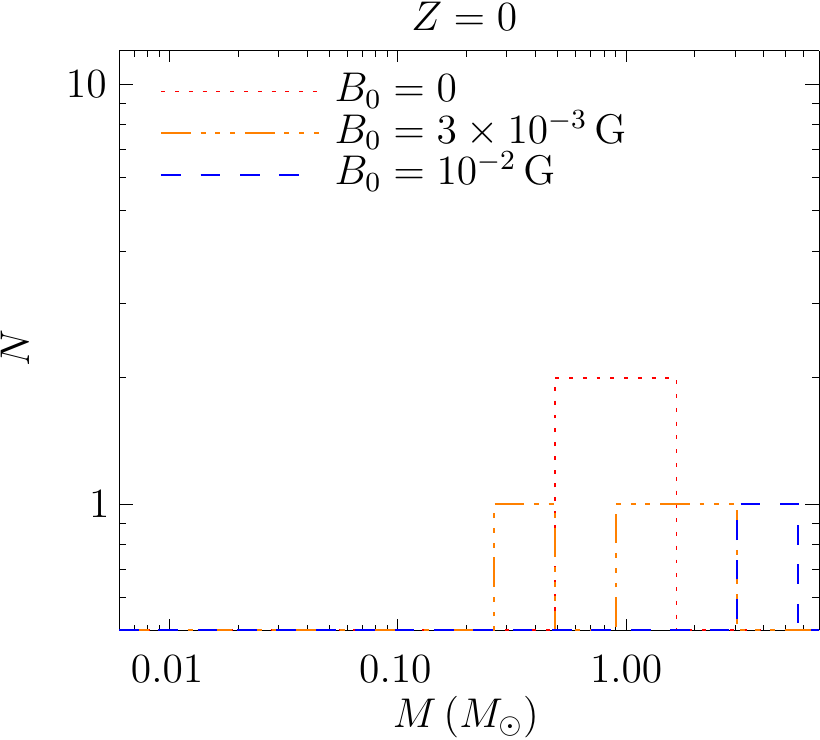}}
\centerline{\includegraphics[height=160pt]{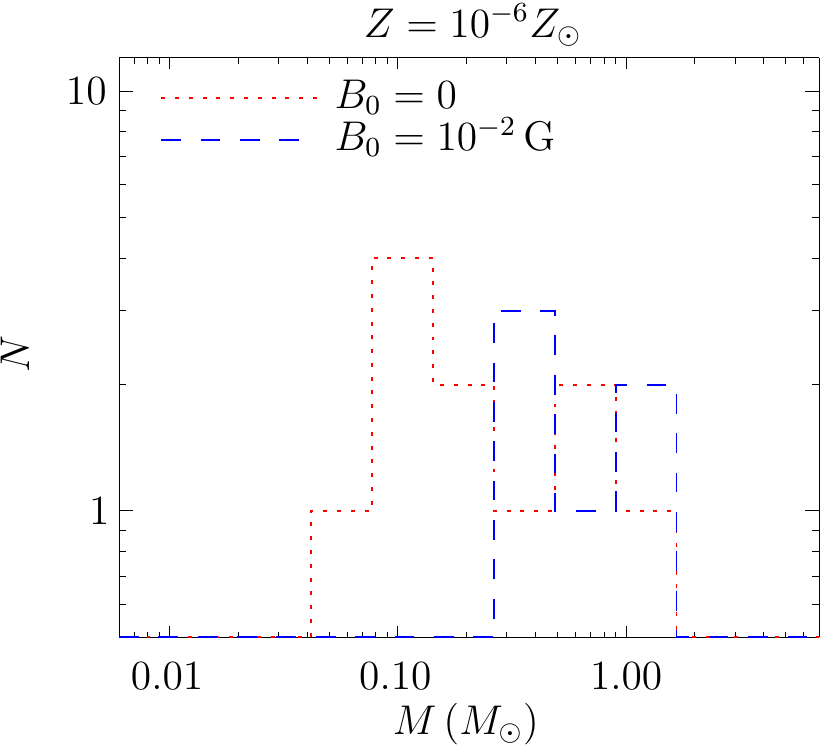}}
\centerline{\includegraphics[height=160pt]{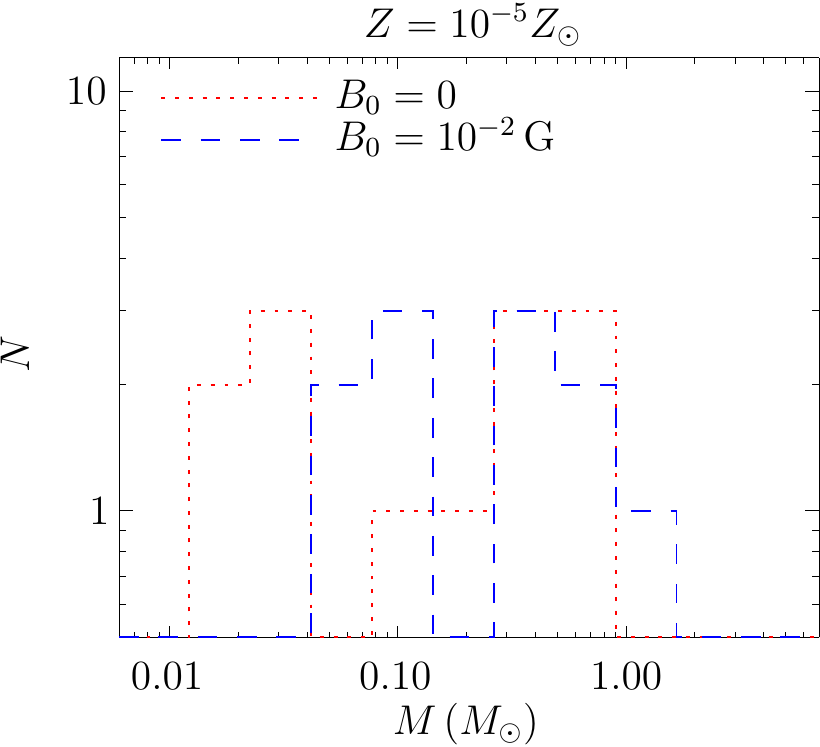}}
\centerline{\includegraphics[height=160pt]{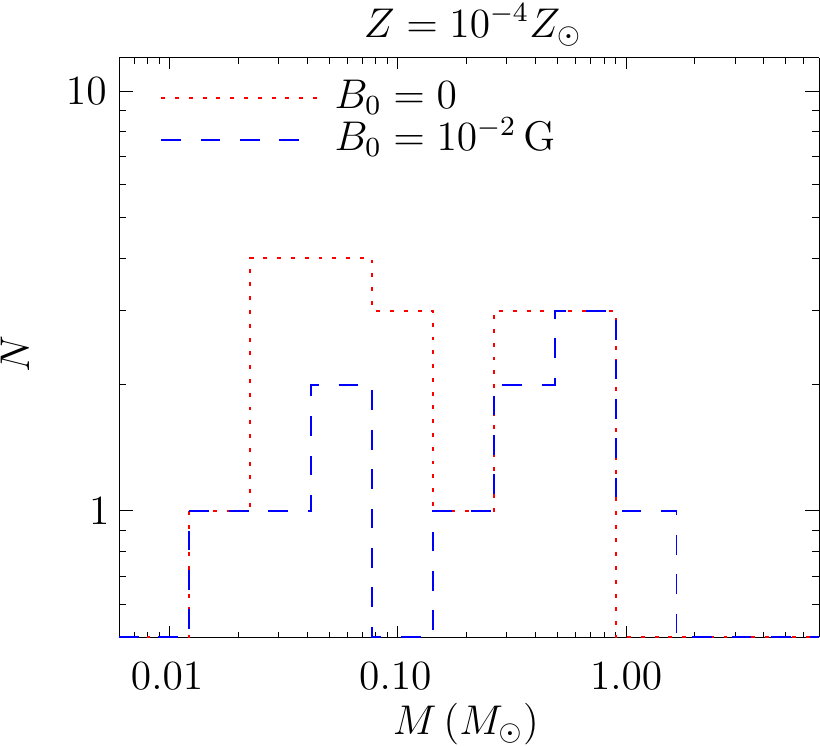}}
\caption{Sink particle mass spectra after approximately $3.75\,M_\odot$ of gas have been accreted onto sinks.}
\label{fig:massspectra}
\end{figure}

The final mass spectra of all simulations are shown in Figure~\ref{fig:massspectra}. The histograms
appear to be flat and show no evidence for a transition in the shape of the initial mass function at
$Z = 10^{-4} Z_\odot$ as observed by \citet{dopcke13}. However, we have accreted $1\,M_\odot$ of gas less in our sink particles
than \citet{dopcke13} and have limited data points in our sample, so that our results are not
statistically significant. What is clear is that in all cases where there is fragmentation a small cluster with a spectrum of
fragment masses develops. These fragments will act as seeds which will grow in mass through accretion, possibly mergers, and
further fragmentation to form a final stellar cluster. The reduction in the number of fragments in the magnetic case suggests
that the final clusters may have fewer stars compared to the non-magnetic case.

\subsection{Timescale Analysis}
\label{sec:timan}

\begin{figure*}
\centerline{\includegraphics[height=160pt]{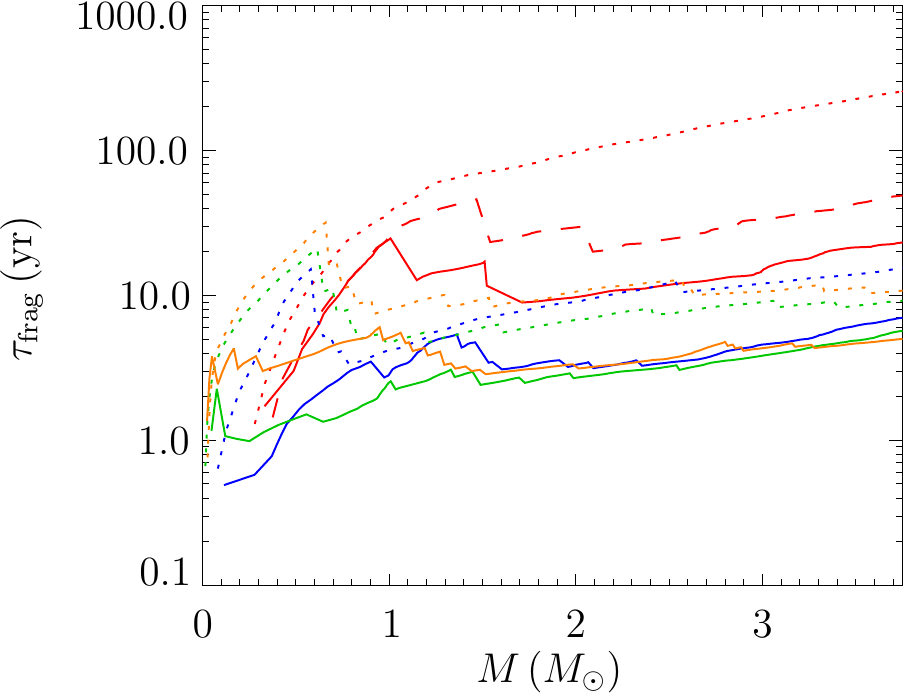}
\includegraphics[height=160pt]{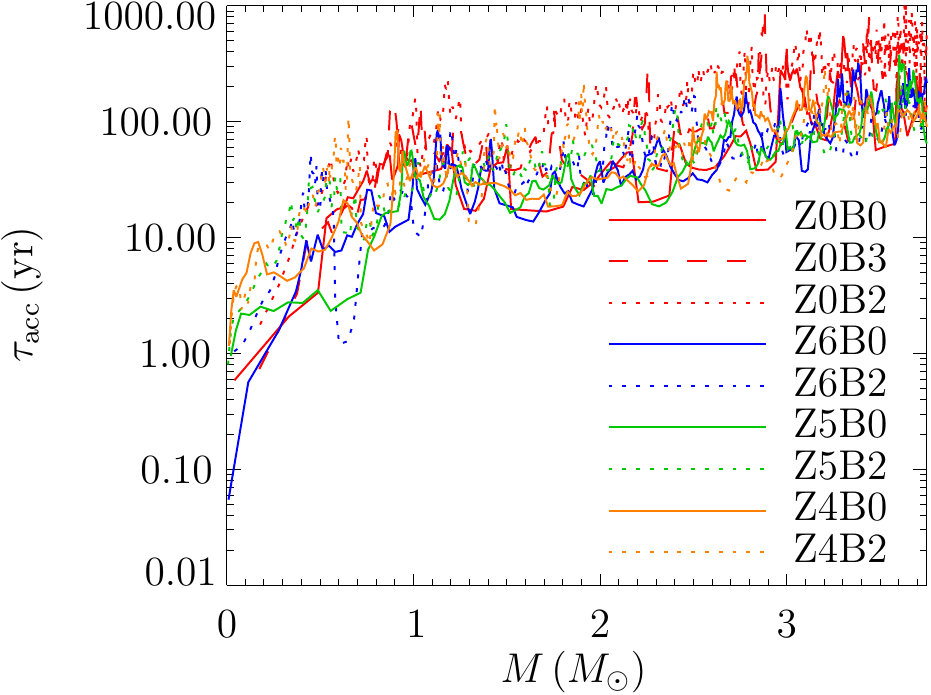}}
\caption{Fragmentation timescale $\tau_\mathrm{frag}$ ({\em left}) and accretion timescale $\tau_\mathrm{acc}$ ({\em right})
as a function of cluster mass for all simulations.}
\label{fig:sinktimescales}
\end{figure*}

We try to understand the differences in the mass spectra with a timescale analysis.
We define a fragmentation timescale
\begin{equation}
\tau_\mathrm{frag}(t) = \frac{1}{N(t)} \sum_{i = 2}^{N(t)} \Delta t_i ,
\end{equation}
with the number of sink particles $N(t)$ at time $t$ and the time difference $\Delta t_i$ between the formation of sink particles
$i$ and $i - 1$. In other words, $\tau_\mathrm{frag}$ is the cluster age divided by the number of stars. The fragmentation timescale
for all simulations as a function of cluster mass is plotted in Figure~\ref{fig:sinktimescales}. Since run Z0B2 only forms a single
sink particle, our figure for $\tau_\mathrm{frag}$ must be considered as a lower limit of the true value.

Most simulations have a fragmentation timescale between $1$ and $30\,$yr, but $\tau_\mathrm{frag}$ for the non-fragmenting run Z0B2
is almost $10$ times larger than this. In this scenario, the disc is particularly stable, as also shown through the analysis of the
support functions in Section~\ref{sec:supfun}. In all cases, $\tau_\mathrm{frag}$ is greater in the magnetic than in the purely hydrodynamic
simulations, typically by about a factor of $3$. In the primordial simulation without magnetic fields, the fragmentation timescale is enhanced
by a factor of at least $10$ compared to the runs with $Z>0$. However, the fragmentation timescale appears to depend very weakly on $Z$ for $Z>0$.
Magnetic fields typically increase it by a factor of $3$, and indeed, the fragmentation timescale in $Z>0$ runs with magnetic fields
is comparable to the $Z=0$ run with no magnetic field.

We compare $\tau_\mathrm{frag}$ with an accretion timescale defined as
\begin{equation}
\tau_\mathrm{acc}(t) = \frac{M(t)}{\dot{M}(t)}
\end{equation}
with the total mass of the cluster $M$ and the total accretion rate onto the cluster $\dot{M}$. The accretion timescale
is shown in Figure~\ref{fig:sinktimescales} as well. Again, magnetised simulations have greater accretion timescales on average than non-magnetic runs.

\begin{figure*}
\centerline{\includegraphics[height=160pt]{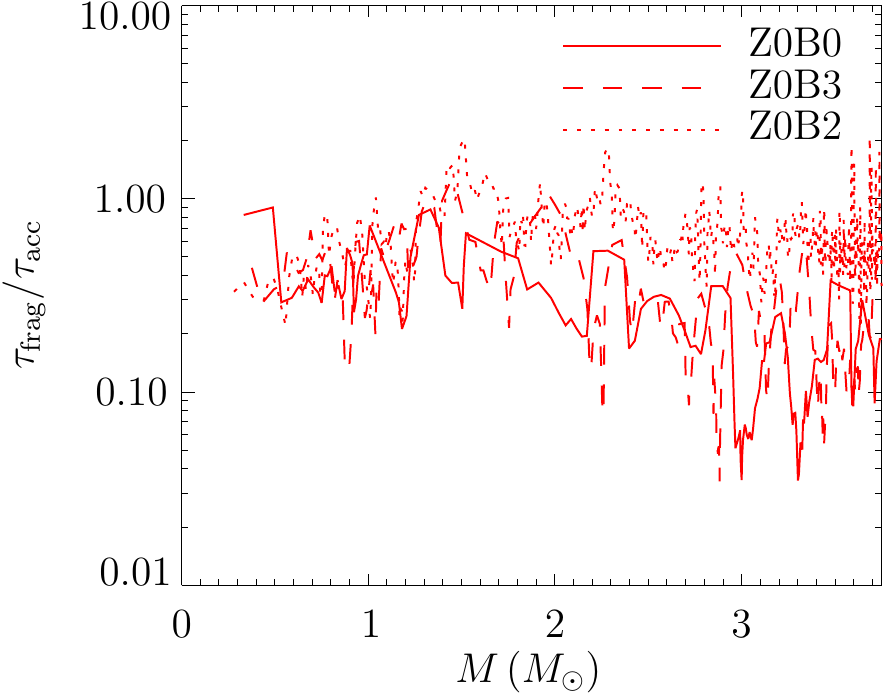}
\includegraphics[height=160pt]{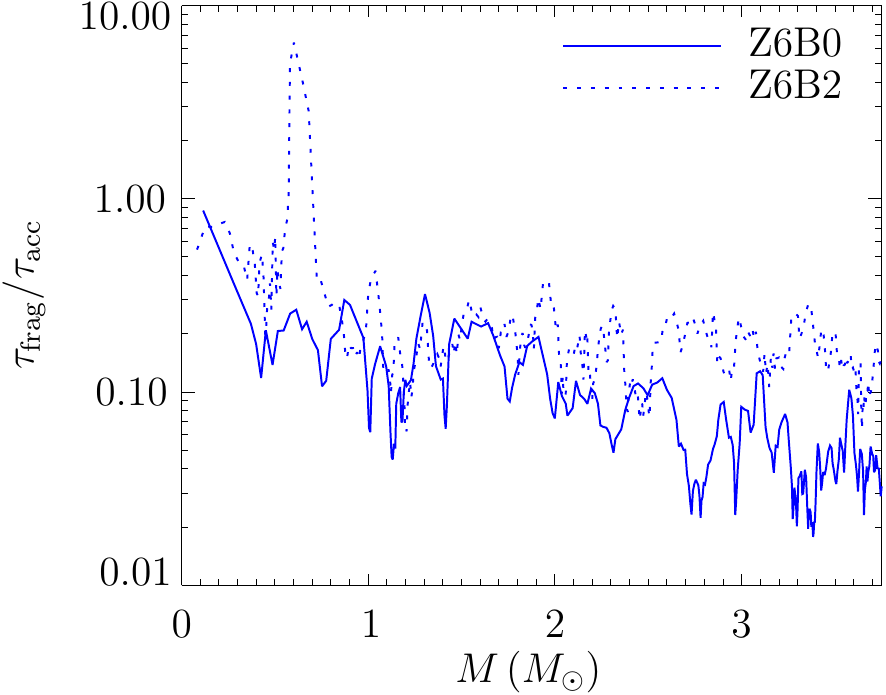}}
\centerline{\includegraphics[height=160pt]{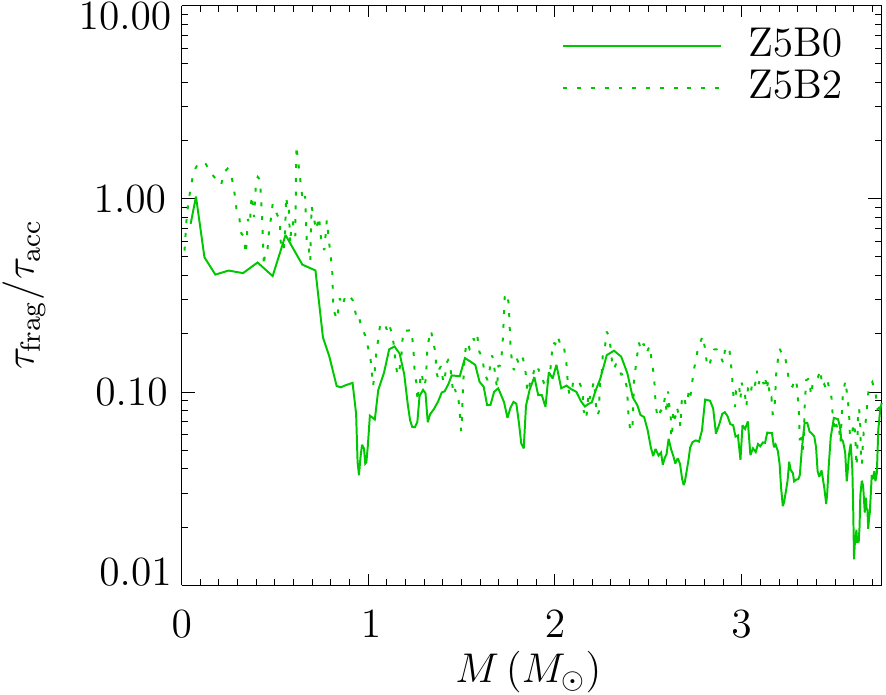}
\includegraphics[height=160pt]{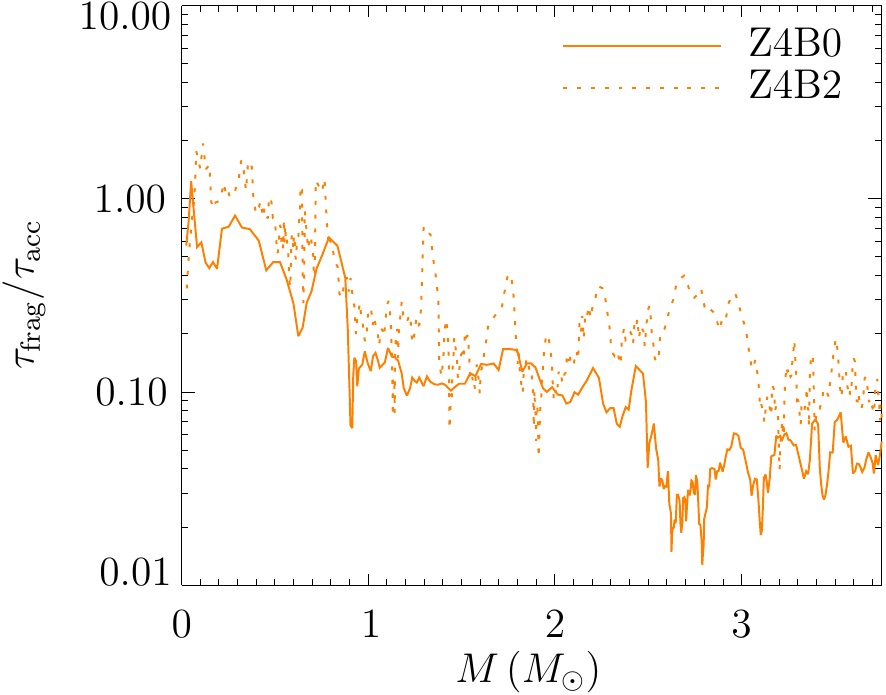}}
\caption{Ratio $\tau_\mathrm{frag} / \tau_\mathrm{acc}$ as a function of cluster mass for all simulations.}
\label{fig:sinktimescalesratios}
\end{figure*}

The ratio of $\tau_\mathrm{frag}$ and $\tau_\mathrm{acc}$ is plotted in Figure~\ref{fig:sinktimescalesratios}. This ratio is smaller
than unity on average and decreases with time. This means that $\tau_\mathrm{acc}$ grows faster than $\tau_\mathrm{frag}$
as the stellar cluster grows. In other words, the stellar system effectively decouples from the supply of gas from the
environment. The central star-forming region is so unstable to gravitational collapse that further accretion
of gas from the halo is not necessary to maintain the star formation activity.
Run Z0B2 is the only simulation for which $\tau_\mathrm{frag} / \tau_\mathrm{acc}$ does
not decrease but oscillates around unity.
The extraordinary stability of the disc in run Z0B2 is consistent with the unusually large support functions
for this simulation (compare Figure~\ref{fig:lambdaZ0B0}).

\section{Discussion}

In this paper, we present a study assessing the relative impact of metals and magnetic fields during low-metallicity star formation.
For this purpose, we consider metallicities ranging from the primordial case ($Z=0$) up to metallicities of $10^{-4}$~$Z_\odot$. The magnetic
field strength has been chosen such that the magnetic energy becomes comparable to the turbulent energy during the formation of the disc, as expected
through the operation of the small-scale dynamo \citep{schleicher10c,schober12}. We further pursue comparison runs with no magnetic fields to quantify
their dynamical impact. In all cases, we follow the collapse until $3.75$~$M_\odot$ of gas are converted into sink particles.

In the runs without magnetic fields, we find that the primordial simulation provides the most stable configuration, where the fragmentation timescale $\tau_\mathrm{frag}$
is enhanced by a factor of $\sim10$ compared to the higher-metallicity runs. For $Z>0$, however, we observe that $\tau_\mathrm{frag}$ depends only
weakly on the actual metallicity. The accretion timescale $\tau_\mathrm{acc}$ of the cluster is slightly enhanced in the primordial case, but not as much as $\tau_\mathrm{frag}$,
and overall $\tau_\mathrm{acc}$ appears to depend only weakly on metallicity. We therefore conclude that the overall accretion is set by the
dynamics of the global collapse and therefore insensitive to the cooling at high densities.

In the presence of magnetic fields, the primordial disc becomes very stable, and $\tau_\mathrm{frag}$ increases by a factor of at least $10$.
Up to the time covered in our simulation ($\sim2\,$kyr), only one sink particle has formed. For $Z>0$, $\tau_\mathrm{frag}$ is still
enhanced by a factor of $3$, but fragmentation is no longer suppressed. We find,
however, a similarity between primordial runs without magnetic fields, and low-metallicity runs with magnetic fields, which have comparable fragmentation timescales.

These results are explored using the thermal, turbulent and magnetic support functions proposed by \citet{schmidt13}. In the absence of
magnetic fields, the thermal pressure yields the dominant positive support for all metallicities explored here, while turbulence yields predominantly
negative contributions due to the presence of compressible motions. On large scales where collapse has already occured, magnetic fields have comparable
negative and positive contributions, while they act as a stabilising agent on the small scale of the protostellar disc. The latter explains why the fragmentation
timescale increases in the presence of magnetic fields.

Finally, we assess the impact of these processes on the sink particle mass functions.
These sink particles represent collapsing fragments, not finished stars.
At the time when we stop our
simulation, i.e. when $3.75$~$M_\odot$ are converted into sinks, we compare the sink particle mass distribution in simulations of different
metallicities and with and without magnetic fields. In the primordial run with a saturated magnetic field, only one single fragment forms, therefore
indicating a potentially top-heavy sink mass function. With decreasing field strength, the number of sinks slightly increased, but remains reduced compared to
the $Z>0$ simulations. A shift towards larger sink masses is therefore expected in the primordial case, which is particularly pronounced in the
presence of magnetic fields.

In the simulations with $Z>0$, the sink mass function appears rather similar regardless of the actual value of $Z$ or the presence of a magnetic field. We note,
however, that the number of sinks is somewhat reduced in all simulations with magnetic fields, and there is a weak trend indicating the formation
of slightly more massive sinks. In order to more strongly constrain the potential influence of metallicity and magnetic fields
on the stellar initial mass function (IMF), we would need to follow the simulations
until a larger number of sinks has formed, therefore improving the statistics for assessing the IMF. However,
one should further explore the impact of different initial conditions, and such simulations would also require
to include additional physics such as radiative feedback \citep[e.g.][]{petersetal10a,petersetal10b,petersetal10c,petersetal11a,smith11,stacyetal12}.
In particular, the data for our barotropic equation of state was extracted from the simulations of \citet{dopcke13} before a disc was formed,
which probably results in an overestimate of the temperature and too stable discs. On the other hand, our equation of state does not include stellar
feedback, which would heat the disc by an unknown amount, stabilising the disc again.
Which of these two competing processes dominates is at present unclear.
Nevertheless, our results indicate a similarity between metal enrichment and magnetic fields in terms of the fragmentation behaviour,
which needs to be explored further in future studies.

\section*{Acknowledgements}

We thank David Collins for technical support and Gustavo Dopcke, Simon Glover and Paul Clark for providing the effective equation of state employed here
as well as for stimulating scientific discussions. We also thank the anonymous referee for useful comments that helped to improve the paper.
T.P. acknowledges financial support through SNF grant 200020\textunderscore 137896 and a Forschungskredit of the University of Z\"{u}rich, grant no. FK-13-112.
D.R.G.S. and W.S. thank the DFG for funding via the Collaborative Research Center (CRC) 963 on {\em Astrophysical Flow Instabilities and Turbulence} (projects A12 and A15). 
D.R.G.S., R.S.K. and R.J.S acknowledge support from the DFG via the SPP 1573 (grants SCHL~1964/1-1, KL~1358/14-1 \& SM~321/1-1)
and via the SFB 881 {\em The Milky Way System} (sub-projects B1, B2, B3).
R.S.K. further acknowledges support from the Baden-W\"{u}rttemberg Foundation via contract research (grant P-LS-SPII/18) as well as from the
European Research Council via the ERC Advanced Grant `STARLIGHT: Formation of the First Stars' (project ID 339177).
We acknowledge computing time at the Leibniz-Rechenzentrum (LRZ) in Garching under project ID h1343, at the
Swiss National Supercomputing Centre (CSCS) under project IDs s364/s417 and at J\"ulich Supercomputing Centre under project ID HHD14.
The FLASH code was in part developed by the DOE-supported Alliances Center for Astrophysical Thermonuclear Flashes (ASCI) at the University of Chicago.
The data was partly analysed with the yt code \citep{turketal11}.

\appendix

\section{Magnetic Field and Density Structure for simulations with $Z > 0$}
\label{app:magden}

\begin{figure*}
\centerline{\includegraphics[width=350pt]{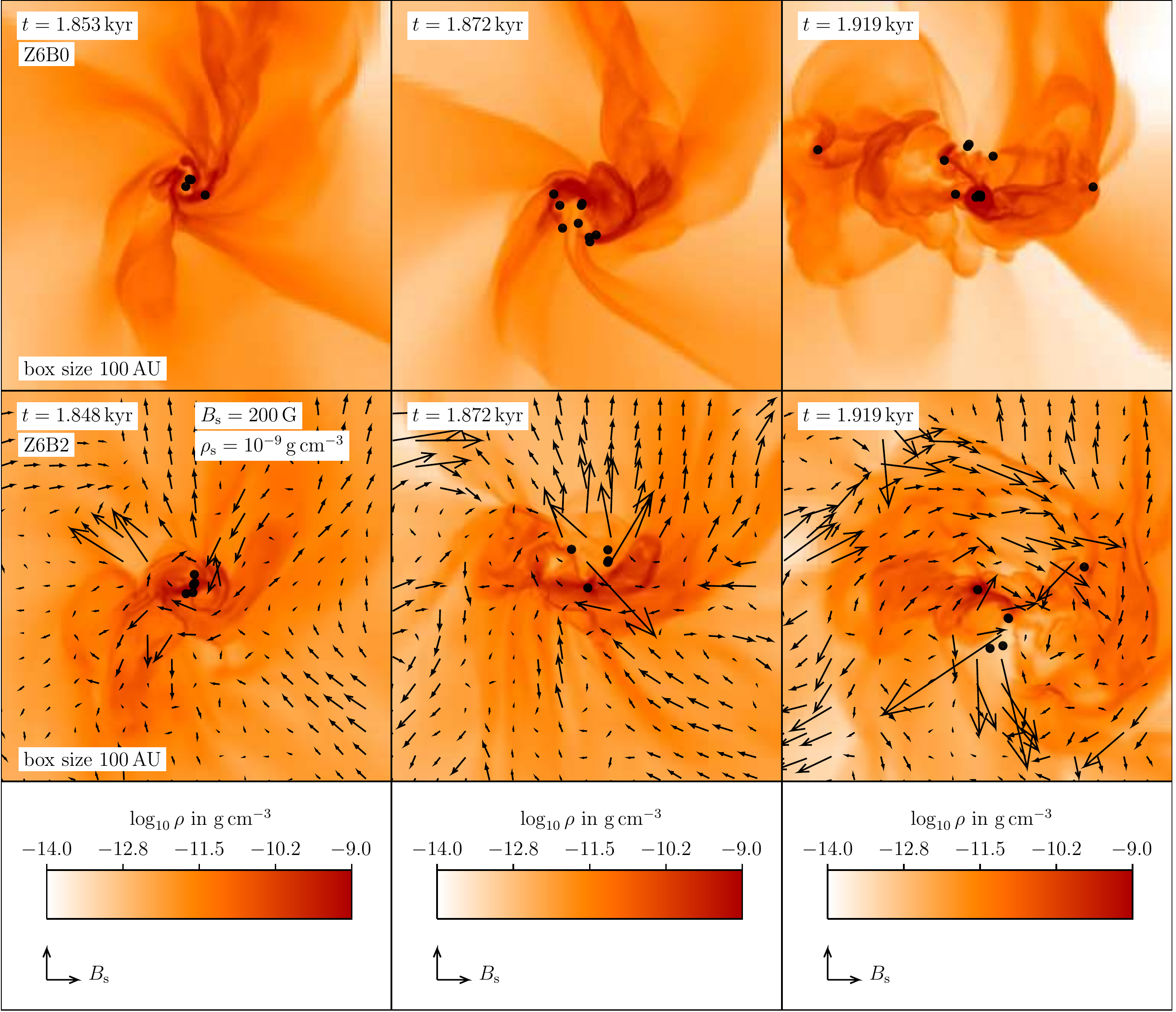}}
\vspace{5pt}
\caption{Magnetic field and density structure for the runs Z6B0 and Z6B2 as a function of time. Rows are different initial magnetic field strengths
($B_0 = 0, 10^{-2}\,$G, from top to bottom), columns are different times (time advances from left to right). The snapshots show
the simulations when the cluster masses have reached $1$, $2$ and $3.75\,M_\odot$, respectively.
The magnetic field vectors have been rescaled for plotting by $(\rho/\rho_\mathrm{s})^{2/3}$, and a field strength of $B_\mathrm{s}$ at a density
of $\rho_\mathrm{s}$ corresponds to an arrow of the length given in the legend. Black dots represent sink particles.}
\label{fig:Z6}
\end{figure*}

\begin{figure*}
\centerline{\includegraphics[width=350pt]{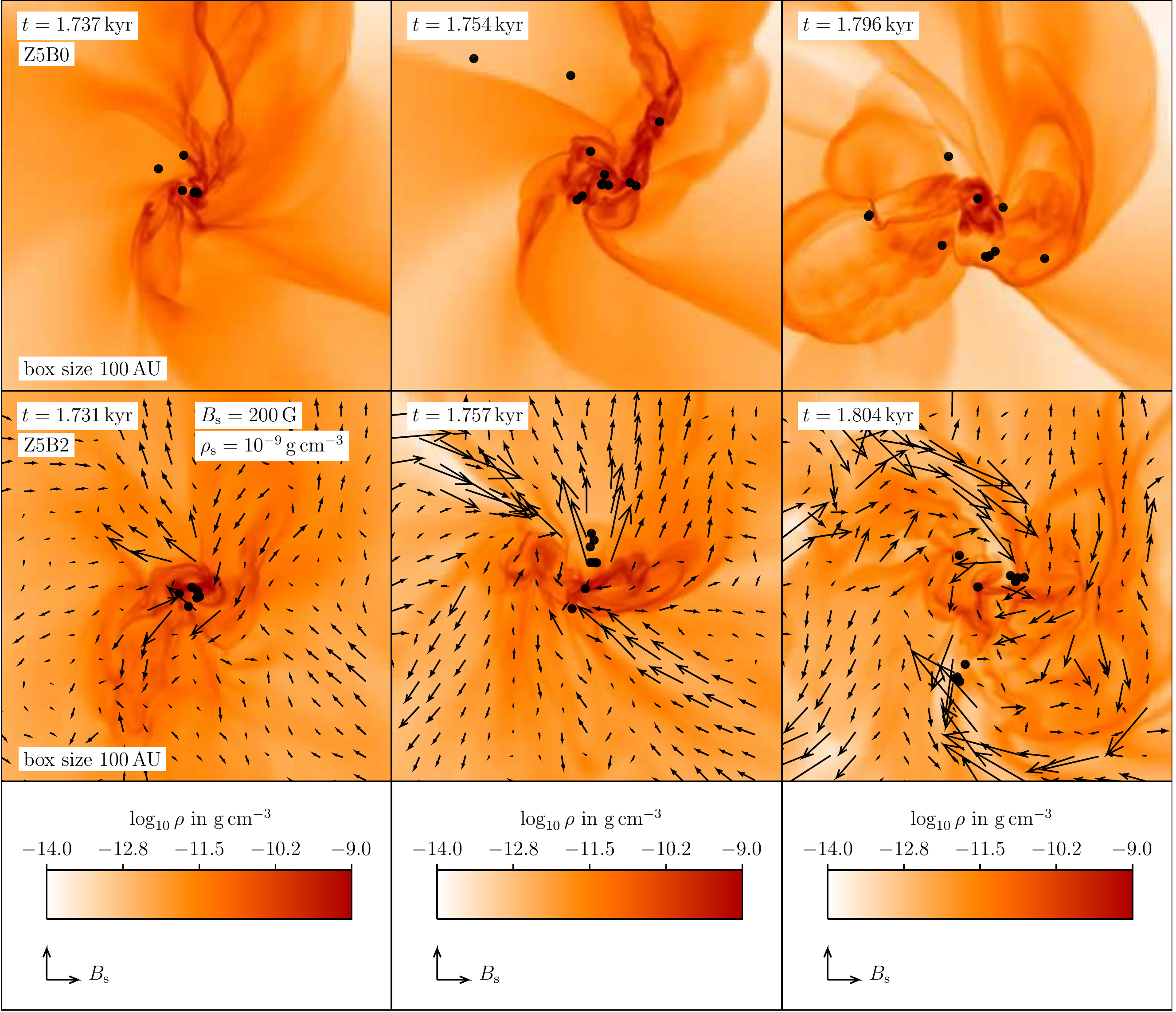}}
\vspace{5pt}
\caption{Magnetic field and density structure for the runs Z5B0 and Z5B2 as a function of time. Rows are different initial magnetic field strengths
($B_0 = 0, 10^{-2}\,$G, from top to bottom), columns are different times (time advances from left to right). The snapshots show
the simulations when the cluster masses have reached $1$, $2$ and $3.75\,M_\odot$, respectively.
The magnetic field vectors have been rescaled for plotting by $(\rho/\rho_\mathrm{s})^{2/3}$, and a field strength of $B_\mathrm{s}$ at a density
of $\rho_\mathrm{s}$ corresponds to an arrow of the length given in the legend. Black dots represent sink particles.}
\label{fig:Z5}
\end{figure*}

\begin{figure*}
\centerline{\includegraphics[width=350pt]{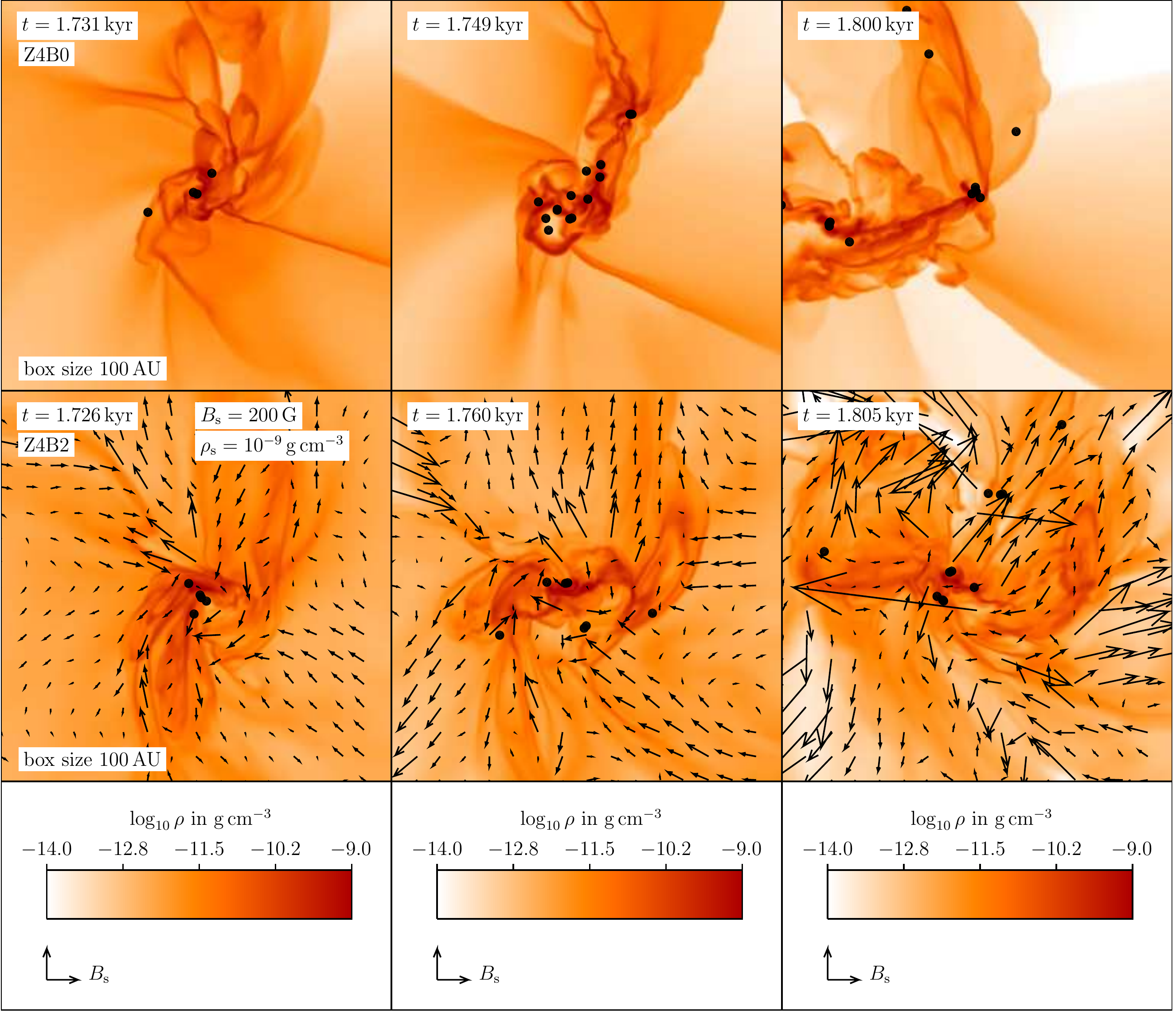}}
\vspace{5pt}
\caption{Magnetic field and density structure for the runs Z4B0 and Z4B2 as a function of time. Rows are different initial magnetic field strengths
($B_0 = 0, 10^{-2}\,$G, from top to bottom), columns are different times (time advances from left to right). The snapshots show
the simulations when the cluster masses have reached $1$, $2$ and $3.75\,M_\odot$, respectively.
The magnetic field vectors have been rescaled for plotting by $(\rho/\rho_\mathrm{s})^{2/3}$, and a field strength of $B_\mathrm{s}$ at a density
of $\rho_\mathrm{s}$ corresponds to an arrow of the length given in the legend. Black dots represent sink particles.}
\label{fig:Z4}
\end{figure*}

In this Appendix, we show the density structures and magnetic field vectors for the simulations with $Z = 10^{-6} Z_\odot$ (Figure~\ref{fig:Z6}), 
$Z = 10^{-5} Z_\odot$ (Figure~\ref{fig:Z5}) and $Z = 10^{-4} Z_\odot$ (Figure~\ref{fig:Z4}) as the sink particles form. The structues look very similar
by and large, but the filaments seem to be more pronounced than in the completely metal-free case $Z = 0$ (Figure~\ref{fig:Z0}).

\section{Support Functions for Simulations with $Z > 0$}
\label{app:supp}

\begin{figure*}
\centerline{\includegraphics[height=140pt]{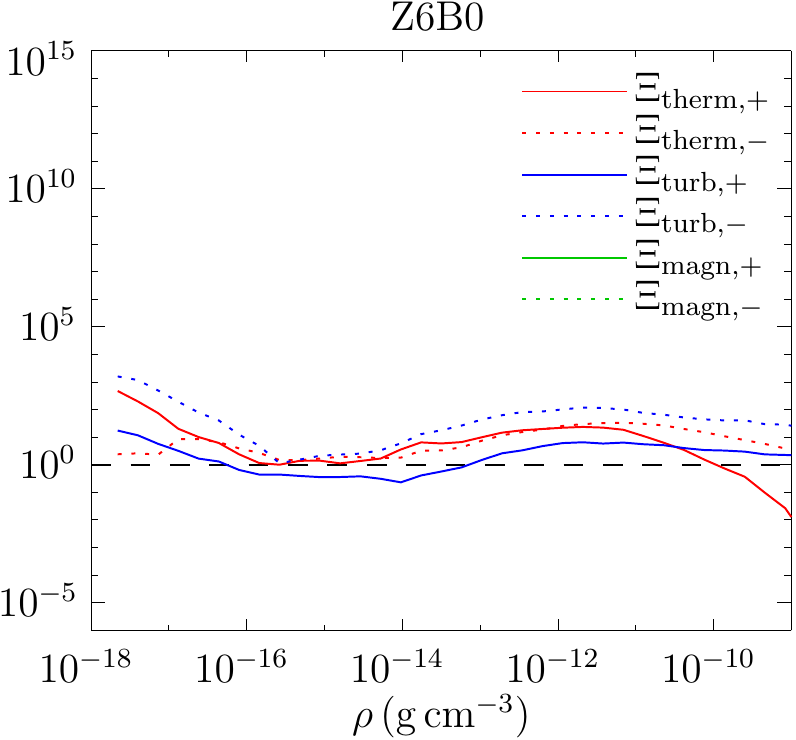}
\includegraphics[height=140pt]{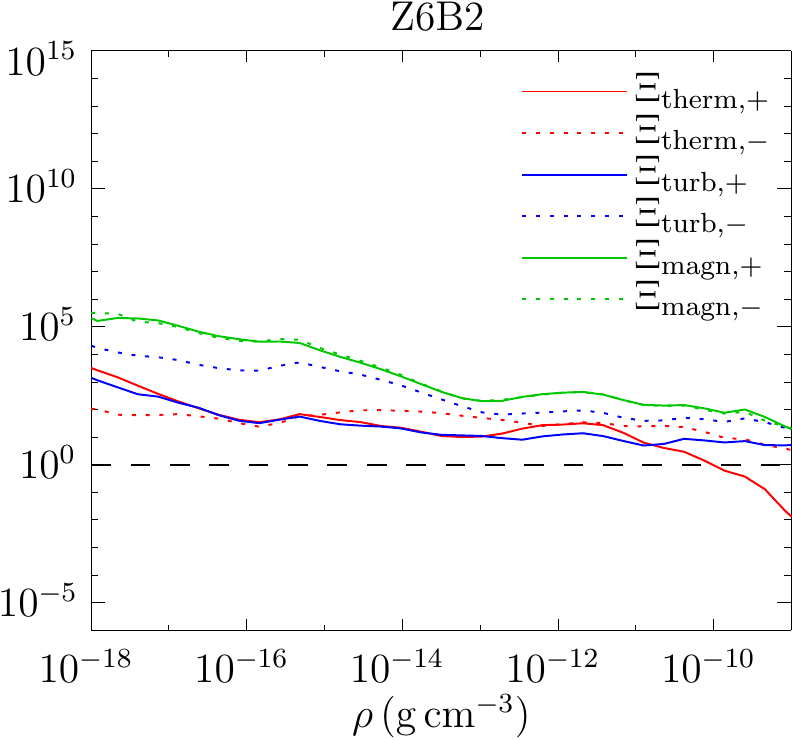}}
\caption{Support functions as a function of density after accretion of $3.75\,M_\odot$ for runs Z6B0 and Z6B2.}
\label{fig:lambdaZ6B0}
\end{figure*}

\begin{figure*}
\centerline{\includegraphics[height=140pt]{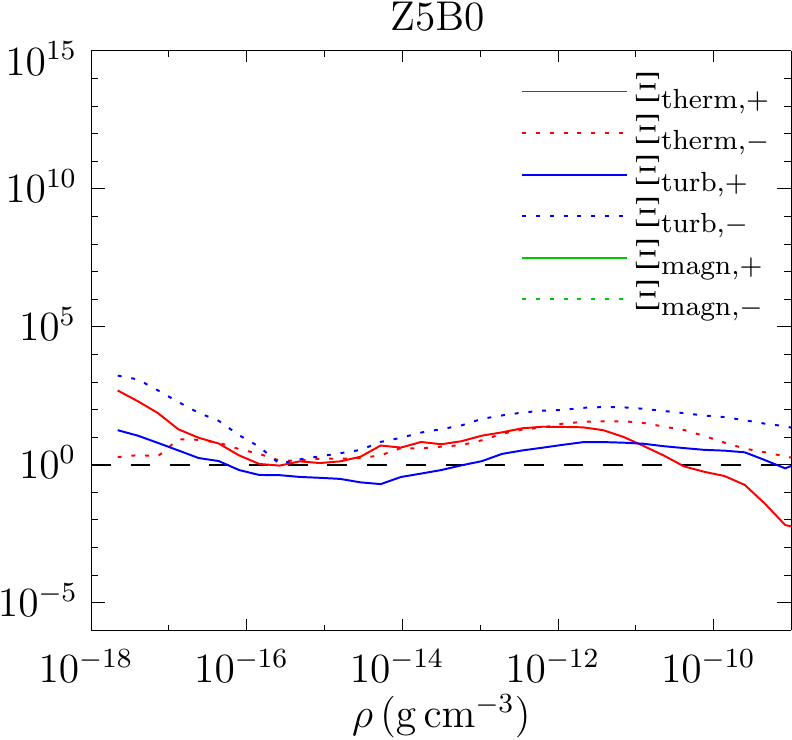}
\includegraphics[height=140pt]{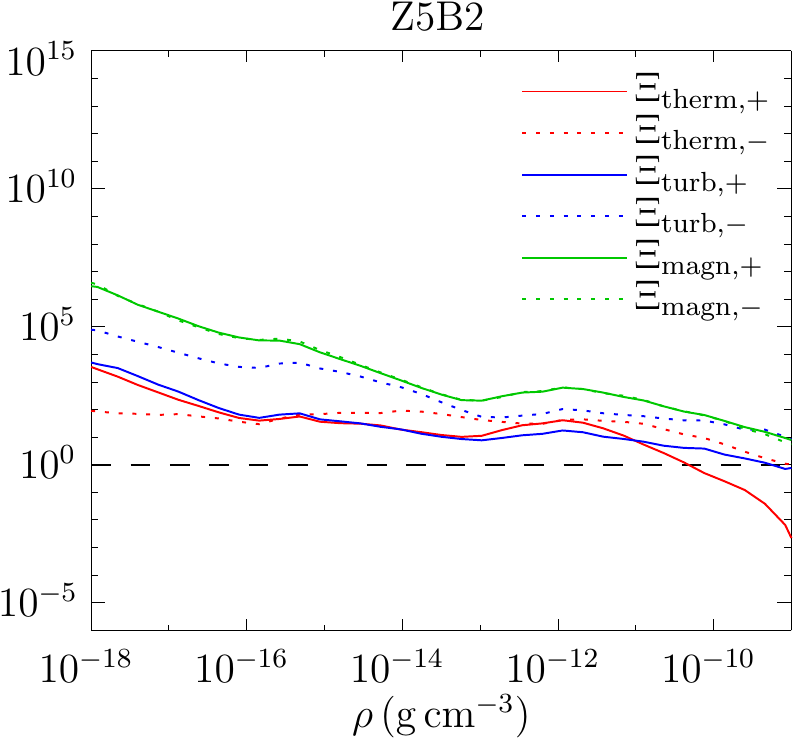}}
\caption{Support functions as a function of density after accretion of $3.75\,M_\odot$ for runs Z5B0 and Z5B2.}
\label{fig:lambdaZ5B0}
\end{figure*}

\begin{figure*}
\centerline{\includegraphics[height=140pt]{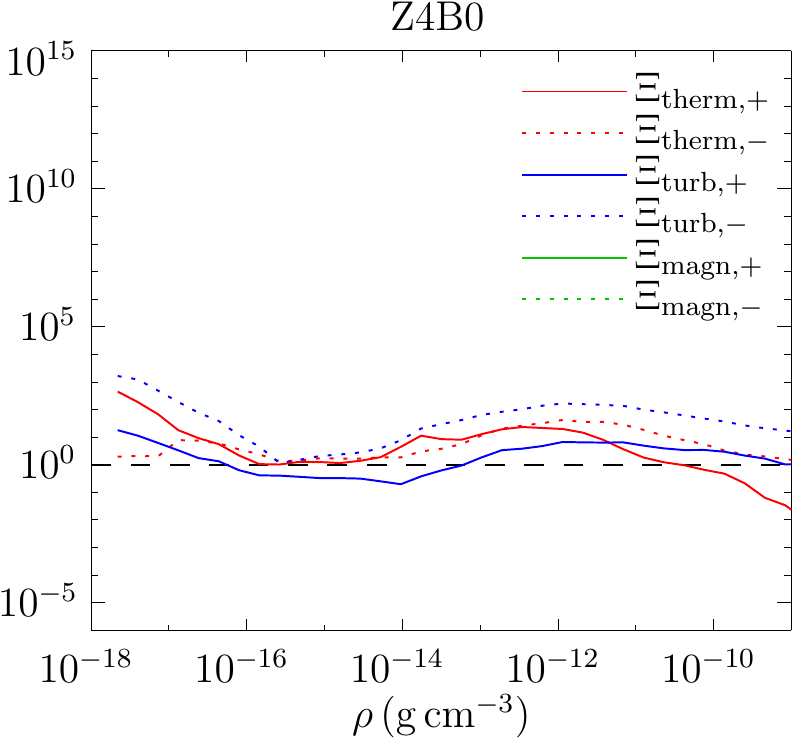}
\includegraphics[height=140pt]{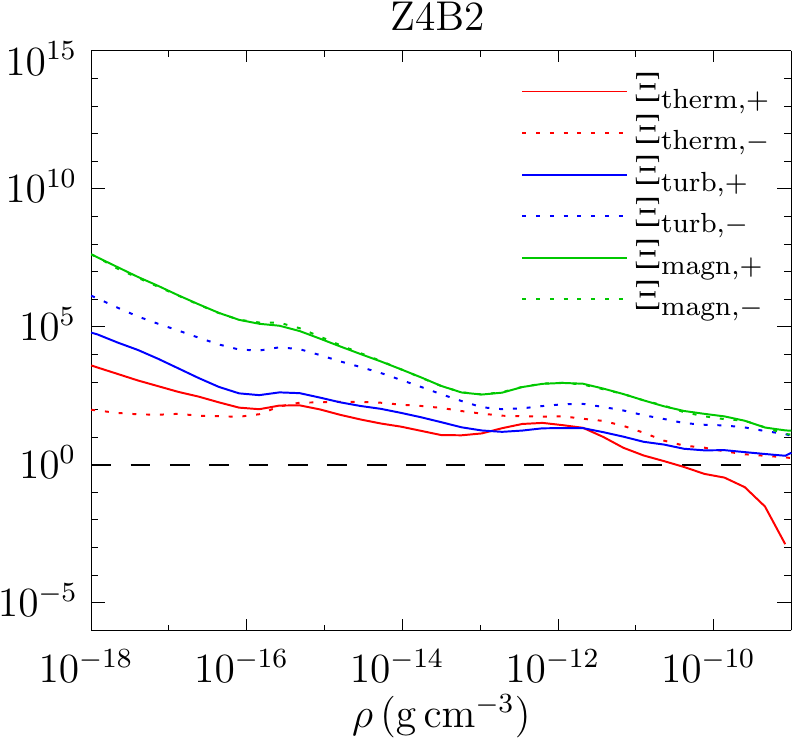}}
\caption{Support functions as a function of density after accretion of $3.75\,M_\odot$ for runs Z4B0 and Z4B2.}
\label{fig:lambdaZ4B0}
\end{figure*}

Here we show plots of $\Xi_\mathrm{therm}$, $\Xi_\mathrm{turb}$ and $\Xi_\mathrm{magn}$ for the runs with  $Z = 10^{-6} Z_\odot$ (Figure~\ref{fig:lambdaZ6B0}),
$Z = 10^{-5} Z_\odot$ (Figure~\ref{fig:lambdaZ5B0}) and $Z = 10^{-4} Z_\odot$ (Figure~\ref{fig:lambdaZ4B0}). There is no strong variation with
metallicity in general. However, the simulation Z0B2 (Figure~\ref{fig:lambdaZ0B0}) has significantly larger support functions than all runs
with $Z > 0$, which explains the reduced fragmentation observed in run Z0B2.

\end{document}